\documentclass{article}

\usepackage{arxiv}

\usepackage[utf8]{inputenc} 
\usepackage[T1]{fontenc}    
\usepackage{hyperref}       
\usepackage{url}            
\usepackage{booktabs}       
\usepackage{amsfonts}       
\usepackage{nicefrac}       
\usepackage{microtype}      
\usepackage{lipsum}
\usepackage{graphicx}
\graphicspath{ {./images/} }

\title{Forecasting Japanese elections: A nonlinear machine-learning approach}

\author{
 Sota Kato\thanks{Corresponding author. Emails: skato@glocom.ac.jp, sotakatoj@gmail.com} \\
  International University of Japan\\
  \And
  Xuan Luo \\
  The Tokyo Foundation\\
  \And
 Budrul Ahsan \\
  IBM Japan\\
  \And
  Asahi Obata \\
  Rice University \\
    \And
  Takafumi Nakanishi\\
  Tokyo University of Technology\\  
}

\begin{document}
\maketitle
\begin{abstract}
Despite Japan being one of the world's largest advanced democracies, the development of election forecasting models for its national elections remains limited. This study introduces nonlinear machine-learning forecasting models, based on decision tree and ensemble learning methods, for predicting the outcomes of Japanese lower-house elections. To assess the methodological benefits of our approach, we replicated the theoretical framework and dataset of Lewis-Beck and Tien's (LBT) foundational statistical forecasting model for Japanese elections. Our models demonstrated moderately but consistently improved predictive accuracy compared to LBT's model in both in-sample and out-of-sample evaluations, suggesting that nonlinear algorithms offer an alternative approach to classical linear methods in capturing complex electoral dynamics. This study represents one of the earlier applications of nonlinear machine-learning techniques to single-country election forecasting. It offers a replicable framework that, when combined with the country-specific electoral theories of other nations, may enhance the predictive performance of forecasting models in broader national contexts.
\end{abstract}


\section{Introduction}
Multiple approaches have been proposed for forecasting elections in democracies worldwide. They can be divided into three principal categories \cite{lewisbeck2011election}: polls, election prediction markets (e.g., \cite{arrow2008promise}), and modeling.\footnote{Bélanger \& Trotter \cite{belanger2017econometric} further add citizen forecasting and social-media-based forecasting as typical approaches to electoral forecasting.} This study uses the third approach, modeling, to forecast Japanese lower-house elections.\footnote{Our approach replicates the approach followed in LBT's forecasting model, which synthesizes poll results. Thus, our models can also be classified as a synthetic model \cite{lewisbeck2014us, lewisbeck2015forecasting}.} Lewis-Beck and Tien (LBT) \cite{lewisbeck2012japanese} pioneered the introduction of a rigorous forecasting model for Japanese lower-house elections based on the modeling approach. Unfortunately, despite Japan being one of the world's major advanced democracies, this foundational work has seen little subsequent research.\footnote{Several attempts have been made to forecast Japanese elections using different approaches. Umeda \cite{umeda2023aggregating} aggregated poll-based media assessments, publicized a few days before the election, to predict 2017 lower-house electoral results at the district and national levels. Nasuno et al. \cite{nasuno2015twitter} used Twitter data to predict the results of 2013 upper-house election.}

The modeling approach for electoral forecasting combines substantive and methodological theories \cite{lewisbeck2011election}. Standard theories of electoral behavior are typically used for the former and regression theory for the latter, as exemplified by the seminal work of LBT's forecasting model of Japanese elections. This study focuses on improving the latter---the methodological theory of electoral forecasting. Hence, we believe that the models developed in this study, when combined with the substantive election theories of other countries, can also serve as a replicable methodological basis for developing country-specific electoral forecasting models.

The election forecasting literature identifies accuracy, lead time, parsimony, and replicability as the standard criteria for evaluating forecasts \cite{lewisbeck2005election, belanger2017econometric}. While we consider all these criteria---ensuring replicability via open code and maintaining the parsimony and lead time of LBT's benchmark model---our primary objective in this study is to improve accuracy. We justify this prioritization by referencing Campbell \cite{campbell2008forecasting}, who asserts that ``the ultimate standard for any forecast... must be its accuracy,'' noting that ``lead time, parsimony, and reproducibility cannot compensate for the inaccuracy of a model's forecasts.'' Similarly, Lewis-Beck \cite{lewisbeck2005election} constructs a model quality index in which the accuracy component is weighted three times heavier than parsimony or reproducibility, reflecting its ``special importance.''

Instead of employing the standard linear regression theory used in LBT's model, we propose an alternative methodological approach using machine-learning algorithms, namely, decision tree (DT) and ensemble-learning methods. As predictive modeling focuses on generating accurate predictions of new observations rather than hypothesis testing \cite{shmueli2010explain}, using machine-learning algorithms over statistical models for election forecasting is defensible. To comparatively examine how our new methodological approach can improve prediction ability over the linear regression approach used in LBT's model, we replaced only the methodological part of LBT's model with our machine-learning models. We kept the substantive part fixed using the same data and variables as those used by LBT for Japanese elections.

One possible reason for the underdevelopment of Japanese electoral forecasting models is the endemic scarcity of observations in single-country forecasts. Only 28 lower-house elections have been held in Japan since the introduction of universal suffrage after World War II, as of mid-2024.\footnote{We included the last two elections held under the Constitution of the Empire of Japan, as they were conducted with universal suffrage.} This small sample size problem poses acute challenges for highly parameterized forecasting models, as it often leads to overfitting of the models \cite{linzer2014future, mitchell1997machine}.

Simple linear models such as those used by LBT to forecast country-specific elections can, to some extent, alleviate the overfitting problem. Indeed, these models often outperform more complex ones in terms of forecasting accuracy \cite{petropoulos2022forecasting, makridakis2018statistical}. However, simple linear models can sometimes lead to underfitting when the underlying relationships are complex and nonlinear \cite{montgomery2018tree}.

Cross-national electoral forecasting leverages larger sample sizes to develop more complex models using machine-learning techniques (e.g., \cite{kennedy2017improving}). However, as the ``no-free-lunch'' theorem suggests \cite{wolpert1995no},\footnote{The no-free-lunch theorem states that ``if an algorithm performs well on a certain class of problems then it necessarily pays for that with degraded performance on the set of all remaining problems'' \cite{wolpert1997no}. Therefore, according to the theory, a universal forecasting model should have limitations when compared with a country-specific forecasting model in forecasting the country's elections.} incorporating country-specific institutional aspects of elections can substantially enhance forecasting accuracy \cite{lewisbeckforthcoming}. For example, LBT's model accounts for features of Japanese electoral institutions to improve its performance. Cross-national forecasting models are limited in capturing such country-specific factors.

The main challenges in this study were thus twofold: 1) introducing nonlinear forecasting models to mitigate the underfitting problem that often appears in simple linear models, and at the same time, 2) limiting the overfitting problem that often arises in nonlinear models trained on small samples. We adopted the DT method as the primary learning algorithm, as it has proven effective for building accurate nonlinear prediction models in the social sciences \cite{montgomery2018tree}. To address overfitting and improve predictive accuracy, we further applied ensemble learning algorithms to develop DT-ensemble models.

We evaluated the performance of our DT-ensemble models against LBT's model, which served as the benchmark. We also compared our models with linear ensemble models to assess the added predictive value of introducing nonlinearity. The DT-ensemble models showed encouraging results, yielding moderately but consistently improved accuracy over both LBT's model and the linear ensemble models in both in-sample and out-of-sample forecasts.

We assumed the same substantive relationships between the independent variables and electoral results as in LBT's model. Thus, we argue that the improved predictive performance of our models can largely be attributed to differences in the methodological approach: our models employed nonlinear DT-based ensembles, whereas LBT's model used a linear statistical model. We also argue that, by combining our models' methodological approach with the substantive theory of countries beyond Japan, our models could contribute to improving other country-specific electoral forecasting models.

\section{Data and models}
\subsection{Data}
\label{sec:data}
To systematically compare our models with LBT's forecasting model, we used the same dataset as LBT: Japanese lower-house election results since the establishment of the Liberal Democratic Party (LDP) in November 1955. The LDP has been the dominant party in Japan and has continuously held office since its establishment, except during 1993--1996 and 2009--2012. We added data for three of the four lower-house elections (i.e., the 2014, 2017, and 2021 elections) held after LBT's paper was published, and used the resulting dataset to train and evaluate both our models and LBT's model.\footnote{We also compared the performance of LBT's model and our models only using the data used in LBT's paper, which do not include 2014 and 2017 elections. The results were not substantially different from those including elections after the publication of LBT's paper (see Appendix Table~\ref{tab:appendix_a1_actual}). Our non-linear models outperformed the LBT benchmark model in both in-sample and out-of-sample forecasting.} We excluded the remaining 2012 election for reasons explained later in this section.

The outcome (dependent variable) to be forecast is, as in LBT's model, the LDP seat occupancy rate (\textit{LDP ratio}). LBT define the \textit{LDP ratio} as the number of seats won by the LDP in a lower-house election divided by the total number of seats in the lower house. For elections held after the comprehensive electoral reform of 1993, the number of seats includes seats won by the LDP in both single-member districts and proportional representation constituencies.

We also used the same three independent variables as in LBT's model in our study. The first is the real gross domestic product (GDP) growth rate (\textit{GDP}). As LBT pointed out, past electoral studies have often found significant correlations between key pre-election economic indicators---such as the GDP growth rate---and the incumbent party's vote share \cite{fair1978effect, inoguchi1981explaining, lewisbeck1984forecasting, hirano2007henyo, hirano2012seiken}. Economic indicators have thus been, along with political approval rates discussed below, central variables in political forecasting models \cite{lewisbeckforthcoming, belanger2017econometric}. Replicating LBT's model, we used the real GDP growth rate from the calendar year prior to each election.

The second variable is the cabinet approval rate (\textit{PM approval}). A positive correlation between approval rates and electoral results is logically intuitive and has been empirically supported by several studies \cite{sigelman1979presidential, lewisbeck1984forecasting}. We followed LBT's model and included the cabinet approval rate surveyed by Jiji Press one month prior to each election.

The third variable is the number of days between two consecutive elections (\textit{Days}). After World War II, Japanese prime ministers have concurrently served as the leaders of the incumbent party and have held the authority to strategically call early elections before their terms expire in the lower house \cite{kato2021time, kayser2005who}. An earlier election implies an advantageous situation for the prime minister and the incumbent party, because if the situation is unfavorable for them, the prime minister can simply wait for conditions to improve until the term expires (four years maximum).\footnote{Note that under the parliamentary system where the Prime Minister can usually call an early election, the time-factor plays different role in electoral forecasting compared with the presidential system (as for presidential forecasting models including time-factor, see, for example \cite{abramowitz1988improved, abramowitz2008time}).}

Among the 22 lower-house elections held between 1958 and mid-2024, we excluded the 1958 election, in accordance with LBT's study, and the 2012 election from our analysis. The 1958 election was excluded because the independent variable \textit{Days} could not be appropriately measured: the LDP was established after the previous election in early 1955. We also excluded the 2012 election because it was the only election during the period that was called by a non-LDP party (the Democratic Party of Japan). Hence, the dependent variable \textit{LDP ratio} does not indicate the incumbent party's seat share for the 2012 election. Table~\ref{tab:descriptive_stats} summarizes the descriptive statistics of the data used in this study. Each variable includes 20 observations corresponding to the 20 elections analyzed.

\begin{table}[htbp]
\centering
\caption{Descriptive statistics of variables}
\label{tab:descriptive_stats}
\begin{tabular}{lcccc}
\toprule
\textbf{Variable} & \textbf{Mean} & \textbf{Standard Deviation} & \textbf{Minimum} & \textbf{Maximum} \\ 
\midrule
\textit{LDP ratio} (\%) & 53.15 & 8.80 & 24.80 & 63.40 \\
\textit{GDP} (growth \%) & 3.83 & 4.11 & -4.40 & 11.91 \\
\textit{PM approval} (\%) & 36.32 & 9.39 & 16.30 & 54.80 \\
\textit{Days} & 1098.45 & 297.14 & 259.00 & 1470.00 \\
\bottomrule
\end{tabular}
\end{table}
In addition, as LBT also implicitly noted, we considered whether to exclude the 2009 election, which stands out as a statistical outlier. As visually demonstrated in the following section, this election deviates significantly from general trends. While several substantive arguments could be made for excluding the 2009 election as an outlier, none of them are decisive. Therefore, in line with LBT, we chose to retain it in our forecasting analysis.\footnote{Nevertheless, we also evaluated the results using a dataset that excluded the 2009 election, a potential outlier, as a robustness check. Our model outperformed the LBT benchmark model more substantially in this case (see Appendix Table~\ref{tab:appendix_a2_actual}).}

\subsection{Models}
\subsubsection{Model development}
\label{subsubsec:model_dev}

\begin{figure}
  \centering
  \includegraphics[width=0.9\textwidth]{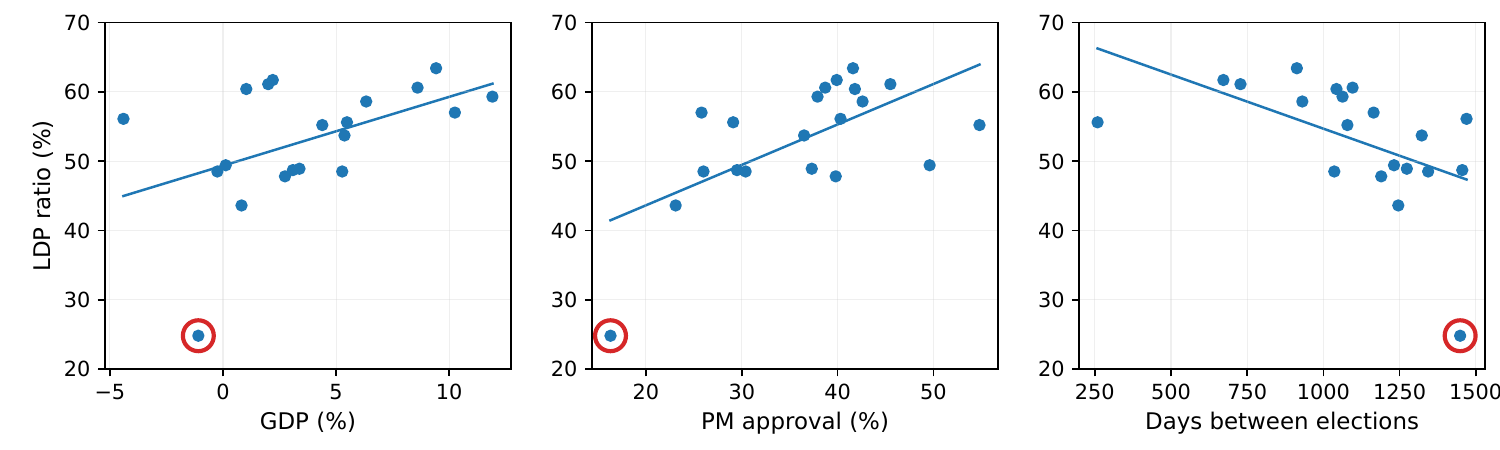}
  \caption{Scatter plot of explanatory variables. LDP: Liberal Democratic Party.}
  \label{fig:fig1}
\end{figure}

Before developing the electoral forecasting models, we plotted each of the three independent variables---\textit{GDP}, \textit{Days}, and \textit{PM approval}---against the outcome of forecasting (\textit{LDP ratio}) to intuitively grasp the data distribution (Figure~\ref{fig:fig1}). Figure~\ref{fig:fig1} suggests that the true relationships between the independent variables and the outcome may be nonlinear, though visual inspection alone is not definitive. It also indicates that, as previously noted, the 2009 election (circled in red) appears to be a statistical outlier.

Beyond visual inspection, the specification of a nonlinear model is grounded in three complementary strands of political science literature. First, regarding institutional mechanics, Taagepera \cite{taagepera1973seats, taagepera1986reformulating} demonstrated that the translation of votes into seats follows a power law (a generalization of the ``Cube Law''), which is a structurally nonlinear relationship. Specifically, Taagepera \cite{taagepera1986reformulating} showed that in Japan's single non-transferable vote (SNTV) system under multi-member districts (MMDs)---prior to the 1994 electoral reform---the seat bonus for large parties accelerates nonlinearly beyond specific vote thresholds. To the extent that the translation of votes into seats follows this power law as Taagepera demonstrated, the functional relationship between cabinet approval (which influences votes) and seat share is expected to be nonlinear as well. More precisely, cabinet approval ratings shape vote shares, and the translation of vote shares into seat shares follows Taagepera's power law; the compound function linking cabinet approval to seat share is therefore nonlinear, even if the first stage---from approval to votes---were approximately linear. The 1994 reform replaced SNTV under MMDs with a mixed-member majoritarian (MMM) system combining single-member districts (SMDs) and proportional representation (PR). The SMD component, governed by first-past-the-post (FPTP) rules, injects the ``Cube Law'' nonlinearity into the post-reform vote-seat relationship. A nonlinear approach can thus capture both the distinct nonlinear functions governing the pre-1994 SNTV system and the post-1994 MMM system, as well as the structural shift between them.

Second, theories of asymmetric voter preferences challenge the linear assumption that voters reward economic gains and punish losses equally. The literature on ``grievance asymmetry'' \cite{bloom1975voter, kappe2018asymmetric} suggests that voters penalize incumbents for downturns but do not reward them proportionately for prosperity, a phenomenon grounded in \textit{Prospect Theory}, where ``losses loom larger than gains.'' Furthermore, mass media may amplify this negativity bias by overemphasizing bad economic news \cite{soroka2006good}. Unlike linear models, which enforce a single symmetric coefficient, decision tree algorithms naturally partition the data to capture these nonlinear conditional effects.

Third, political outcomes are often determined by interactive complexity rather than simple additive effects. Beck, King, and Zeng \cite{beck2000improving} argue that complex political processes, such as international conflict, often involve ``massive nonlinear interactive effects'' where outcomes depend on the confluence of multiple causal factors. For example, the impact of economic indicators on seat share may be highly conditional on the timing of the election. As Montgomery and Olivella \cite{montgomery2018tree} argue, tree-based models are an appropriate theoretical choice for such data because they automatically partition the covariate space to uncover these conditional structures without imposing a rigid functional form.

Given these theoretical considerations and the patterns observed in Figure~\ref{fig:fig1}, simple linear regression models may not adequately capture the underlying relationships, potentially leading to underfitting. Thus, we built nonlinear models combining DT and ensemble methods and compared their performance with that of LBT's benchmark linear model. We also constructed linear ensemble models using the same ensemble methods applied in the DT ensemble models to assess how the nonlinearity of our DT ensemble models improved predictive performance.

Regarding the primary learning algorithm employed for modeling, we used DTs instead of alternatives such as support vector machines or neural networks. DTs can be a valuable algorithm for social scientists, especially for predictions rather than theory testing \cite{montgomery2018tree}. They require fewer assumptions and can incorporate flexible functional forms. Unlike linear models which require the researcher to pre-specify a parametric form (e.g., log-linear or polynomial), DTs can automatically approximate complex piecewise-constant functions from the data without imposing a rigid structure \cite{kaufman2019improving, montgomery2018tree}. Another advantage of DTs over other nonlinear learning algorithms (e.g., neural networks or support vector machine), whose input-output relations are often black-boxed, is their interpretability \cite{james2021introduction}.\footnote{Some scholars even insist that the explainability of the DT makes it the most preferable machine learning algorithm \cite{seni2010ensemble}.} While classical linear regression offers superior interpretability through parametric coefficients, DTs provide transparent rule-based structures and feature importance measures. This interpretability makes DT particularly suitable for political science forecasting models since political scientists traditionally have valued substantive interpretations of independent variables.

Next, we combined multiple DT models using ensemble methods to address potential overfitting problems \cite{dietterich2000ensemble, kaufman2019improving, montgomery2018tree, james2021introduction}. Combining DT models through ensemble methods can also enhance their predictive capacity and help mitigate underfitting problems. We used the DT-based ensemble methods implemented in the \textit{scikit-learn} package of Python (Version 3.6) for our models.\footnote{See \url{https://scikit-learn.org/stable/tutorial/machine_learning_map/} for details.}

The following sections briefly describe our learning algorithms, which combine DT methods (Section~\ref{subsubsec:dt}) and ensemble learning methods (Section~\ref{subsubsec:ensemb_learn_method}). We then explain how we used the training dataset to train and validate our models for performance evaluation (Section~\ref{subsec:model_training}).

\subsubsection{DT}
\label{subsubsec:dt}
A DT is a non-parametric, nonlinear supervised learning method that can perform regression and classification tasks with minimal data preprocessing \cite{hastie2009elements, james2021introduction}. It can handle both qualitative and quantitative data and effectively capture nonlinear relations between independent and dependent variables.

A DT partitions the dataset into smaller subsets during the learning process and recursively develops a tree structure. A non-leaf node performs a split test on the explanatory variables per iteration, dividing the corresponding dataset into subsets. This iterative splitting continues until a predefined stopping criterion is met.

DT-based models are often prone to overfitting; that is, they tend to perform well on training data but poorly on test data, resulting in low predictive accuracy. Furthermore, unconstrained tree models can be susceptible to ``fishing expeditions'' if used to search for predictors across a large number of candidate variables. To prevent this, we deliberately restricted our model inputs to the three pre-specified, theoretically grounded variables used in the LBT benchmark. We employed the DT methodology solely to allow the data to determine functional forms among these theory-specified variables---capturing the nonlinearities and interactions justified in Section~\ref{subsubsec:model_dev}---rather than as a tool for variable discovery. To address the inherent instability of single trees, we combined multiple tree models using ensemble methods to further mitigate the overfitting problem.

\subsubsection{Ensemble learning methods}
\label{subsubsec:ensemb_learn_method}

Ensemble learning methods are used to develop models by combining homogeneous and heterogeneous learners \cite{zhou2012ensemble, freund1997decision}. These methods have improved the performance of various models, including DTs, across a wide range of machine-learning tasks \cite{caruana2006empirical, james2021introduction, kaufman2019improving}. They are also known to substantially reduce output variance when each DT is independent \cite{zhou2015two}. Some ensemble methods have been successfully applied to political predictions \cite{montgomery2012improving, montgomery2015calibrating, brair2020forecasting, hare2023measuring}.

Ensemble methods are commonly classified into two groups: 1) sequential and 2) parallel techniques. Sequential techniques iteratively train multiple base models by correcting the errors of the previous model in the sequence. At the end of the process, an aggregated model with reduced bias is developed. Leading sequential techniques include gradient boosting, which we employed in this study, and its variants such as \textit{XGBoost} and \textit{LightGBM}. These methods have recently produced winning solutions in multiple Kaggle forecasting competitions \cite{januschowski2022forecasting}.

In gradient boosting, a model is first trained on the training dataset. The residuals from this model---the difference between the actual and predicted values---are then calculated and used to train a new model. This sequential model-building process continues until the residual loss falls below a predefined threshold or reaches zero.

Parallel ensemble techniques independently train base models such as individual DTs and integrate them to create a single DT ensemble. Bagging \cite{breiman1996bagging} and random forests \cite{breiman2001random} are the leading methods in this category. Bagging (bootstrap aggregation) builds multiple base models using bootstrapped subsets of the data and aggregates their predictions by averaging the outputs of all models in the final ensemble. A random forest further improves bagging by randomly selecting a subset of independent variables at each split to generate diverse base models, which are then combined to create the final ensemble.

As the number of explanatory variables in our electoral forecasting models is small, the predictive performance of ensemble models created by bagging and random forest algorithms is expected to differ little. Therefore, for parallel ensemble techniques, we used only the bagging algorithm for our forecasting models. We also constructed linear ensemble models to comparatively assess the impact of nonlinearity in our DT ensemble models.

\subsection{Model training, validation, and testing}
\label{subsec:model_training}


We trained and validated our models according to the procedure described in this paragraph. Given the small sample size ($N=20$ elections), relying on a single train-test split could yield results that are highly sensitive to the specific partition of the data. To address this and ensure robust performance evaluation, we employed a repeated random sub-sampling validation (also known as Monte Carlo cross-validation) procedure \cite{picard1984cross}.

First, we randomly shuffled the dataset and split it into a training set (75\%) and a test set (25\%). Next, using only the training set, we tuned the hyperparameters of our models via \textit{leave-one-out cross-validation} (LOOCV). This method is particularly well-suited for small datasets as it maximizes the training data available for tuning. The grid search for hyperparameter tuning varied the following parameter groups: 1) the depth of each DT and the minimum number of samples required for a split and for a leaf node; 2) the number of DT models in the ensemble; and 3) the learning rate in the sequential models.

After determining the optimal hyperparameters, we retrained the models on the specific training split and evaluated their predictive accuracy on the held-out test set. In each iteration, we also re-estimated the coefficients of the LBT benchmark model using the same training split, which includes the 2014, 2017, and 2021 elections that were held after the publication of LBT's paper. This ensures a strictly fair comparison where both models are trained on the identical updated dataset under the same conditions. 

We used the \textit{mean absolute error} (MAE) as the primary evaluation metric, while also calculating the \textit{root mean squared error} (RMSE) as a supplementary measure. We prioritized these direct error metrics over goodness-of-fit statistics such as $R^2$. As Shmueli \cite{shmueli2010explain} argues, $R^2$ assesses in-sample explanatory power, from which out-of-sample predictive power cannot be reliably inferred. Furthermore, goodness-of-fit metrics can be misleading in forecasting tasks due to overfitting risks \cite{petropoulos2022forecasting, harrell2015regression}.

To mitigate the bias associated with any single random split, we repeated this entire procedure---shuffling, splitting, tuning, training, and testing---100 times. The final performance metrics reported in this study represent the average results across these 100 independent trials. This approach differs from a one-step-ahead rolling forecast, in which only elections preceding the target election are used for estimation. With only 20 observations, such a temporal split would leave the earliest folds with very few training cases, yielding unstable estimates.

\section{Results of the performance evaluation}

To evaluate our models' performance, we first conducted in-sample evaluations to assess how well our models fit the training data, compared with the LBT benchmark and linear ensemble models (Section~\ref{subsec:in_sample}). We then proceeded to the key analysis of this study---out-of-sample evaluations---to examine the models' ability to forecast electoral outcomes using the test dataset (Section~\ref{subsec:out_of_sample}).

In the first stage of out-of-sample evaluation (Section~\ref{subsubsec:out_of_sample_perf_ev}), we compared the predictive performance of the linear ensemble models with that of the LBT benchmark linear model. This comparison establishes a baseline for evaluating, in the next section, how introducing nonlinearity through DT ensemble models enhances predictive accuracy. In the second stage (Section~\ref{subsubsec:out_of_sample_perf_ev_non_ensem}), we conducted our main experiment, evaluating the performance of our DT ensemble models against that of the LBT benchmark.

\subsection{In-sample performance evaluation}
\label{subsec:in_sample}
\begin{figure}
  \centering
  \includegraphics[width=0.9\textwidth]{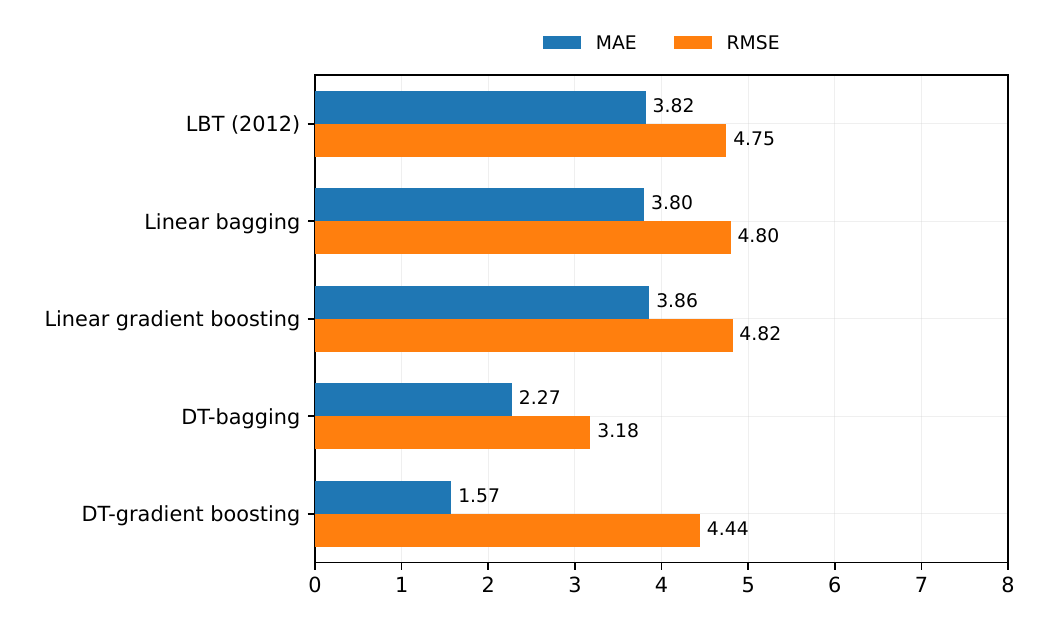}
  \caption{In-sample performance of the DT-based ensemble models. LBT: Lewis-Beck and Tien; DT: decision tree; MAE: mean absolute error; and RMSE: root mean squared error.}
  \label{fig:fig3}
\end{figure}

The results of the in-sample performance evaluation are presented in Figure~\ref{fig:fig3} and Table~\ref{tab:in_sample}. Our DT ensemble models achieved a better fit to the data than both the benchmark and linear ensemble models. \textit{MAE} improved from 3.82 for the benchmark model to 2.27 for the DT-bagging model and to 1.57 for the DT-gradient boosting model.\footnote{As shown in Figure~\ref{fig:fig3}, \textit{MAE} of our DT ensemble models improved more than \textit{RMSE}. This is because \textit{RMSE} is more sensitive to statistical outliers, in this case, the 2009 election.} Similarly, in terms of \textit{RMSE}, our models showed improvements, with the DT-bagging model achieving 3.18 and the DT-gradient boosting model achieving 4.44, compared to 4.75 for the benchmark. These results were anticipated. We reasonably presumed that, for in-sample forecasting, our nonlinear models could address underfitting problems more effectively than simple linear models such as the LBT benchmark model. However, these favorable results may also have been due to the overfitting of our models. In the next section, we present the results of out-of-sample testing to assess whether our models managed to strike an appropriate balance between the underfitting and overfitting.

\begin{table}[htbp]
\centering
\caption{Summary of in-sample performance evaluation results}
\label{tab:in_sample}
\small
\begin{tabular}{lcccccc}
\toprule
 & & \multicolumn{2}{c}{\textbf{Linear ensemble}} & & \multicolumn{2}{c}{\textbf{Nonlinear DT ensemble}} \\
\cmidrule(lr){3-4} \cmidrule(lr){6-7}
 & \textbf{LBT} & \textbf{Bagging} & \textbf{Gradient boosting} & \textbf{DT} & \textbf{Bagging} & \textbf{Gradient boosting} \\
\midrule
\textbf{MAE} & 3.82 & 3.80 & 3.86 & 2.11 & 2.27 & 1.57 \\
 & (0.49) & (0.47) & (0.41) & (0.63) & (0.31) & (0.52) \\
\addlinespace
\textbf{RMSE} & 4.75 & 4.80 & 4.82 & 2.77 & 3.18 & 4.44 \\
 & (0.67) & (0.67) & (0.57) & (0.76) & (0.59) & (1.60) \\
\bottomrule
\multicolumn{7}{l}{\footnotesize Note: Numbers in parentheses represent standard deviations. LBT: Lewis-Beck and Tien;}\\
\multicolumn{7}{l}{\footnotesize DT: decision tree; MAE: mean absolute error; RMSE: root mean squared error.}
\end{tabular}
\end{table}

\subsection{Out-of-sample performance evaluation}
\label{subsec:out_of_sample}
\subsubsection{Out-of-sample performance evaluation: Linear ensemble models}
\label{subsubsec:out_of_sample_perf_ev}

\begin{figure}
  \centering
  \includegraphics[width=0.9\textwidth]{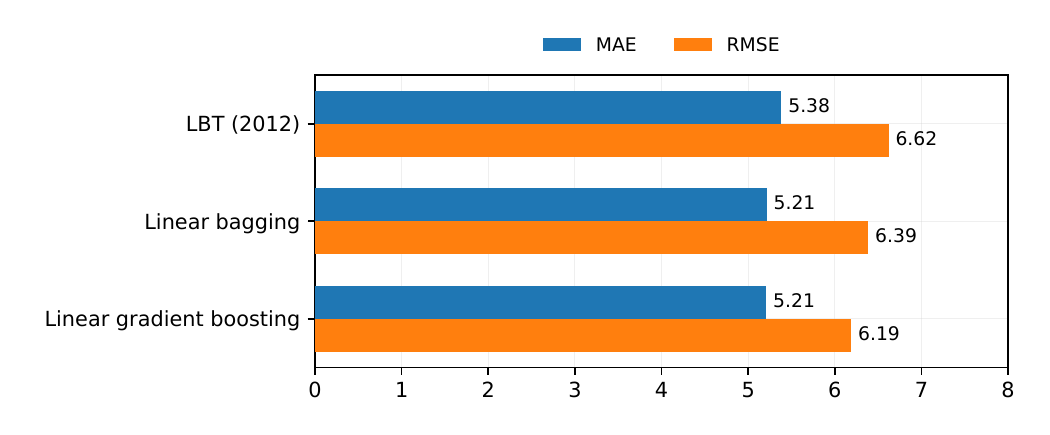}
  \caption{Out-of-sample performance of the linear ensemble models.}
  \label{fig:fig4}
\end{figure}

First, we examined whether forecasting performance could be improved using linear ensemble models. We took this step to precisely assess how introducing nonlinearity to our DT models, which we evaluate in the next section, could improve predictive performance. The gradient boosting and bagging methods described earlier were used for the linear ensemble models. 

The performance evaluation results are shown in Figure~\ref{fig:fig4}. The linear ensemble models scored slightly better than the LBT benchmark model (\textit{MAE} of 5.21 compared to 5.38), but not by a meaningful margin. These results suggest that ensemble methods alone were insufficient to meaningfully improve performance over the benchmark model. In the next section, we assess how introducing nonlinearity through DT-based models could improve performance.

\subsubsection{Out-of-sample performance evaluations: Nonlinear ensemble models}
\label{subsubsec:out_of_sample_perf_ev_non_ensem}

\begin{figure}
  \centering
  \includegraphics[width=0.9\textwidth]{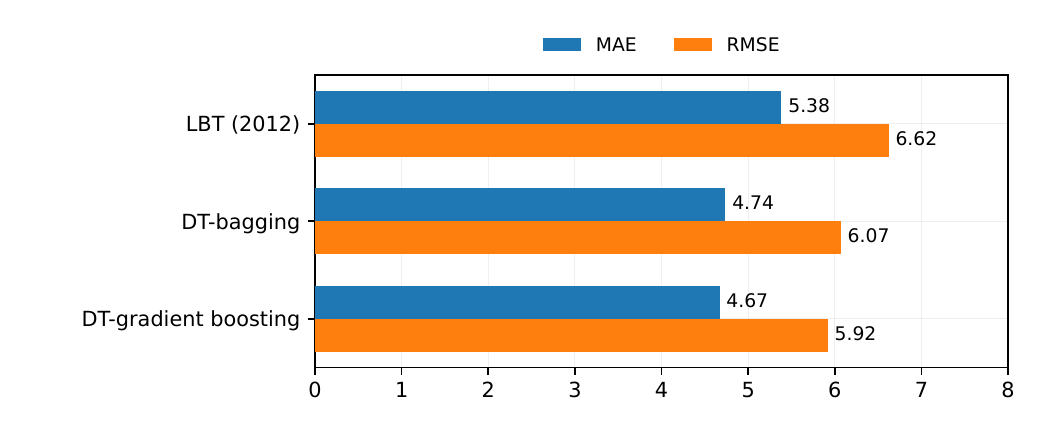}
  \caption{Out-of-sample performance of the DT-based ensemble models.}
  \label{fig:fig5}
\end{figure}
We used the test data to evaluate the generalization performance of our nonlinear, DT-based forecasting models against the LBT benchmark model; the results are shown in Figure~\ref{fig:fig5}. Both our nonlinear ensemble models---DT models combined with bagging and gradient-boosting algorithms---outperformed the LBT benchmark model. The DT-gradient boosting model achieved an \textit{MAE} of 4.67 compared to the LBT benchmark's 5.38, an improvement of 0.71 points (13.2\%). The DT-bagging model achieved an \textit{MAE} of 4.74, an improvement of 0.64 points (11.9\%). While modest in absolute terms, these improvements carry practical significance in the Japanese institutional context. In the House of Representatives, several consequential seat thresholds shape the governing party's legislative power: a simple majority (233 seats), a ``stable majority'' enabling the party to chair all standing committees (244 seats), and an ``absolute stable majority'' enabling the party to pass bills through committees without opposition support (261 seats). Elections are frequently decided near these thresholds, where even a margin of a few seats carries significant political consequences. The 0.7 percentage-point \textit{MAE} improvement, corresponding to approximately 3 seats out of 465, falls within this consequential range.

Regarding the \textit{RMSE}, the DT-gradient boosting model scored 5.92 compared to the LBT benchmark's 6.62, a reduction of 0.70 points (10.6\%) that is nearly identical in magnitude to the 0.71-point (13.2\%) improvement observed in \textit{MAE}. It is worth noting, however, that \textit{RMSE} is known to be highly sensitive to outliers because squaring the errors gives disproportionate weight to large deviations \cite{hyndman2006another}. This property is consequential for our dataset. In the restricted sample covering elections through 2009 (Appendix Table~\ref{tab:appendix_a1_actual}), the \textit{RMSE} values for the two models are identical (5.80)---a parity driven by the 2009 election, in which the LDP suffered a historic defeat that acts as a statistical outlier. When this outlier is excluded from the full dataset (Appendix Table~\ref{tab:appendix_a2_actual}, elections through 2017 excluding 2009), our model outperforms the benchmark by a similar margin on both \textit{MAE} (4.08 vs. 4.56) and \textit{RMSE} (4.79 vs. 5.33), confirming that the model consistently improves predictive accuracy across metrics.

Table~\ref{tab:out_of_sample} summarizes the results of the out-of-sample experiments. Drawing on the results in Tables~\ref{tab:out_of_sample} and \ref{tab:in_sample}, we note that our nonlinear forecasting models predicted Japanese electoral outcomes more accurately than the LBT benchmark model. Our nonlinear models appear to have achieved an appropriate balance between overfitting and underfitting. They also outperformed the linear ensemble models in both in-sample and out-of-sample forecasting, implying that the nonlinearity of our models was an important factor in outperforming the benchmark. Finally, as indicated in Table~\ref{tab:out_of_sample}, our DT ensemble models also predicted electoral outcomes more accurately than a non-ensembled DT model in out-of-sample forecasting.

\begin{table}[htbp]
\centering
\caption{Summary of out-of-sample evaluation performance}
\label{tab:out_of_sample}
\small
\begin{tabular}{lcccccc}
\toprule
 & & \multicolumn{2}{c}{\textbf{Linear ensemble}} & & \multicolumn{2}{c}{\textbf{Nonlinear DT ensemble}} \\
\cmidrule(lr){3-4} \cmidrule(lr){6-7}
 & \textbf{LBT} & \textbf{Bagging} & \textbf{Gradient boosting} & \textbf{DT} & \textbf{Bagging} & \textbf{Gradient boosting} \\
\midrule
\textbf{MAE} & 5.38 & 5.21 & 5.21 & 4.83 & 4.74 & 4.67 \\
 & (2.10) & (1.93) & (2.52) & (2.12) & (1.63) & (1.61) \\
\addlinespace
\textbf{RMSE} & 6.62 & 6.39 & 6.19 & 6.07 & 6.07 & 5.92 \\
 & (2.51) & (2.34) & (2.99) & (2.94) & (2.57) & (2.45) \\
\bottomrule
\multicolumn{7}{l}{\footnotesize Note: Numbers in parentheses represent standard deviations. LBT: Lewis-Beck and Tien;}\\
\multicolumn{7}{l}{\footnotesize DT: decision tree; MAE: mean absolute error; RMSE: root mean squared error.}
\end{tabular}
\end{table}
\begin{figure}
  \centering
  \includegraphics[width=0.9\textwidth]{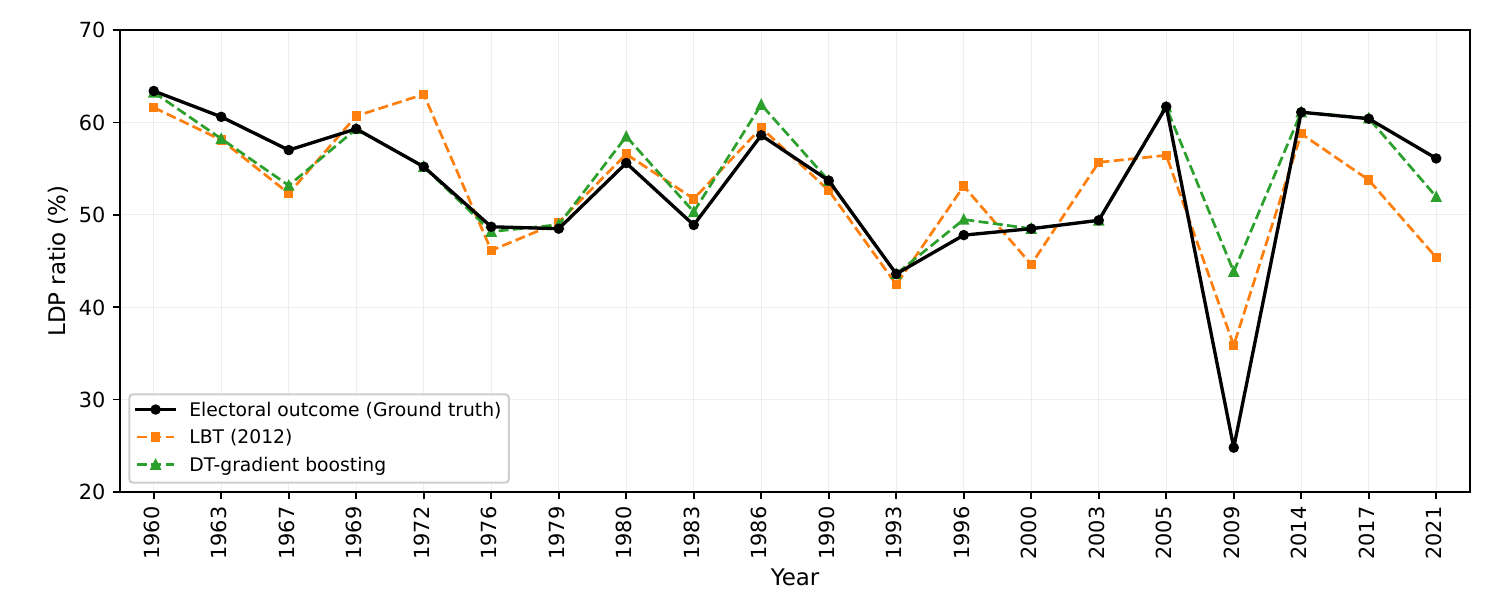}
  \caption{Actual electoral outcomes (ground truth) and predicted values.}
  \label{fig:fig6}
\end{figure}

\begin{figure}
  \centering
  \includegraphics[width=0.9\textwidth]{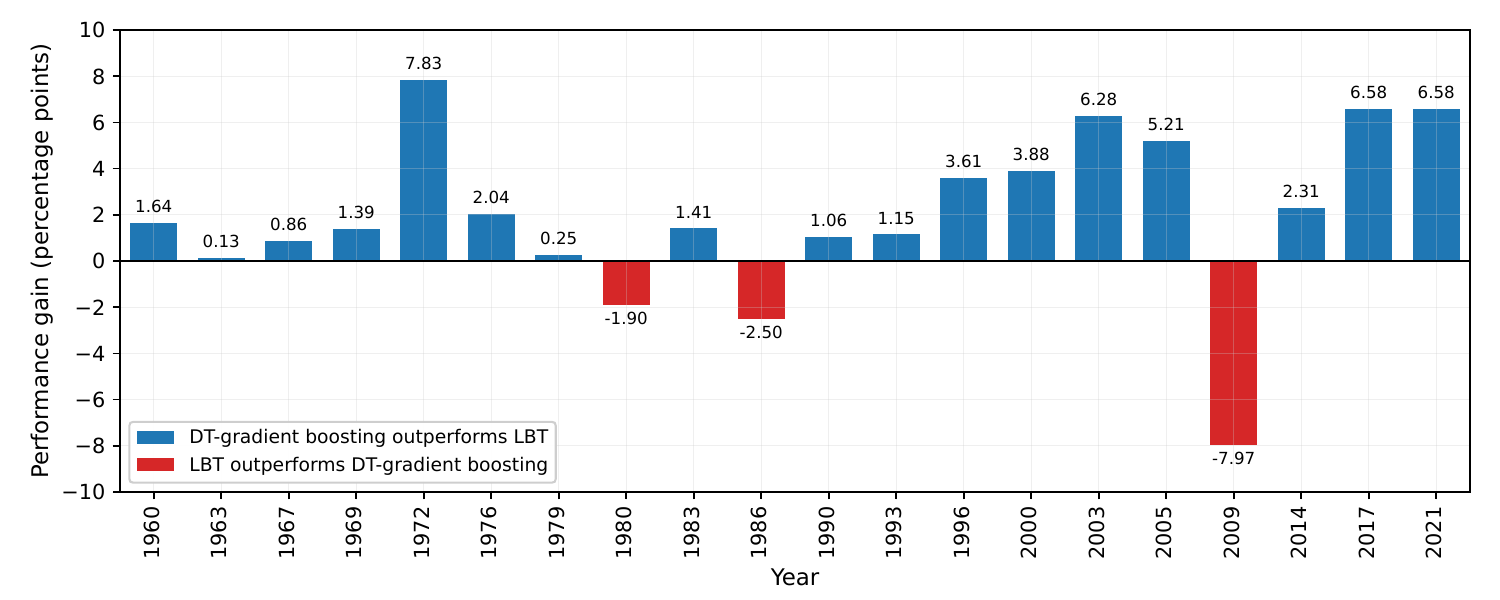}
  \caption{Performance gain over LBT's benchmark model. LBT: Lewis-Beck and Tien.}
  \label{fig:fig7}
\end{figure}

Figure~\ref{fig:fig6} displays the actual outcomes of each election (``ground truth'') and the predicted values from our forecasting model and the LBT benchmark model. Only the DT-gradient boosting model is presented in the figure, as it outperformed the DT-bagging model. Figure~\ref{fig:fig7} shows the performance gain of our model over the LBT benchmark, relative to the actual outcomes (``ground truth''). For example, a gain of 6.58 points in the 2021 election indicates that our model's prediction was 6.58 points closer to the actual LDP ratio than the prediction made by the LBT model.

As a complementary measure of predictive consistency, Figure~\ref{fig:fig6} and Figure~\ref{fig:fig7} demonstrate that our nonlinear model (the DT-gradient boosting model) outperformed the LBT benchmark model in 17 out of 20 elections. This indicates that the model's improvement is robust across time rather than being driven by a few large gains. Excluding elections that were held after the LBT's paper was published, our model still outperformed the LBT's in 14 out of 17 elections (also see Appendix Table~\ref{tab:appendix_a1_actual}). The only election in which our model substantially underperformed relative to the benchmark was the 2009 election, which is considered a statistical outlier. As mentioned earlier, excluding this election as an outlier improved our model's performance relative to the benchmark (\textit{MAE}: 4.08 vs. 4.56; \textit{RMSE}: 4.79 vs. 5.33; see Appendix Table~\ref{tab:appendix_a2_actual}).\footnote{As mentioned, we nevertheless decided not to treat this election as an outlier and included it in our training and test datasets.}

\section{Discussion}

\begin{figure}
  \centering
  \includegraphics[width=0.9\textwidth]{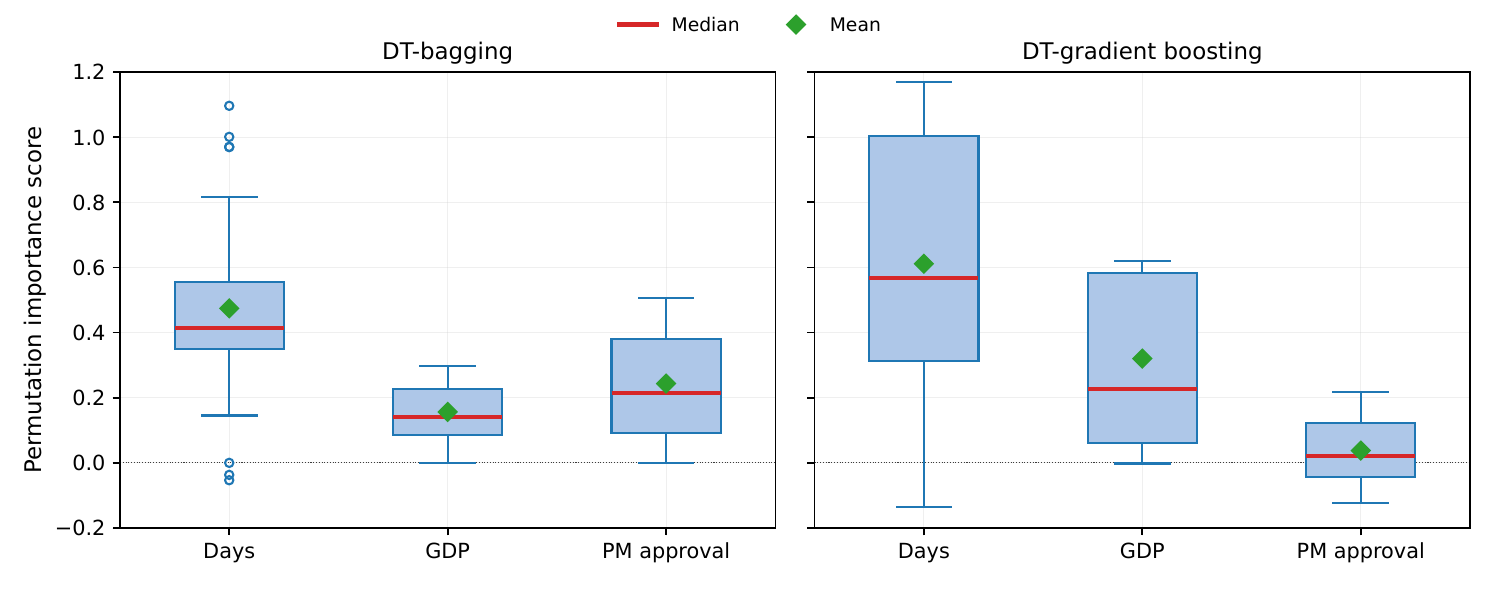}
  \caption{Permutation importance of explanatory variables. Red solid lines indicate the median, and green diamonds indicate the mean.}
  \label{fig:fig8}
\end{figure}

The modeling approach to election forecasting, as noted earlier, combines substantive and methodological theories. In this study, we sought to hold the substantive aspect constant by using the same explanatory variables and the dataset employed in LBT's model, thereby enabling us to focus on the methodological dimension of forecasting. Our models, therefore, still have room for improvement regarding the substantive dimension of Japanese electoral forecasting. In this section, we briefly outline a promising direction for such improvements---drawing on the feature importance measure of our DT-based model---as a potential avenue for future research. We also demonstrate how this measure can enhance the interpretability of electoral forecasting models.

Tree-based models are known to be effective in making predictions when there are numerous candidate predictors whose interrelationships are theoretically unclear \cite{montgomery2018tree}. The feature importance measure is a tool for tree-based models that can be used to identify and prioritize appropriate predictors from candidate features (i.e., independent variables). It typically computes the permutation importance, which measures the decrease in model performance when a feature's values are randomly shuffled, to estimate each feature's relative importance in predicting the outcome~\cite{breiman2001random}.

We computed permutation importance scores for each explanatory variable in our DT-based ensemble models with 30 random shuffles per feature. The results are summarized in Table~\ref{tab:feature_importance_data}, and Figure~\ref{fig:fig8} displays the corresponding box plots. Among the three independent variables, \textit{Days} received the highest importance score in both DT ensemble models, indicating that it contributed the most to accurately predicting electoral outcomes. Regarding the other two variables, \textit{PM approval} scored higher than \textit{GDP} in the bagging model, whereas the reverse was observed in the gradient-boosting model. 

\begin{table}[htbp]
\centering
\caption{Feature importance of the explanatory variables}
\label{tab:feature_importance_data}
\begin{tabular}{lcccccc}
\toprule
 & \multicolumn{2}{c}{\textbf{Days}} & \multicolumn{2}{c}{\textbf{GDP}} & \multicolumn{2}{c}{\textbf{PM approval}} \\
\cmidrule(lr){2-3} \cmidrule(lr){4-5} \cmidrule(lr){6-7}
 & \textbf{Bagging} & \textbf{Grad. boost} & \textbf{Bagging} & \textbf{Grad. boost} & \textbf{Bagging} & \textbf{Grad. boost} \\
\midrule
\textbf{Mean} & 0.474 & 0.611 & 0.156 & 0.320 & 0.243 & 0.038 \\
\bottomrule
\end{tabular}
\end{table}

One of the key advantages of ensemble DT models over other nonlinear machine-learning approaches lies in their ability to identify the substantive variables that enhance forecasting accuracy \cite{kaufman2019improving}. Future research could leverage these feature importance measures, in conjunction with behavioral and institutional analyses of Japanese politics, to refine variable selection and further boost the predictive performance of DT-based models. This modeling strategy could also be extended to other countries by incorporating their respective election theories to develop accurate country-specific forecasting models.

The feature importance measures of tree-based models also offer researchers a means of interpreting these models \cite{breiman1984classification}. One of the limitations of machine-learning models, compared to traditional statistical models, is their lack of interpretability \cite{shmueli2010explain}. Feature importance measures can help mitigate this limitation by identifying the variables that most influence predictions. For instance, it may come as a surprise that \textit{Days} received a higher importance score than more straightforward predictors such as \textit{PM approval} and \textit{GDP}. While this may initially seem counterintuitive, it aligns with recent theoretical work on endogenous election timing in political science \cite{smith2004election, kayser2005who, kato2013valuable}. The high importance score for \textit{Days} lends empirical support to these theories, suggesting that the prime minister's strategic decision of when to call an election---a variable capturing institutional power---is a stronger predictor of electoral outcomes than standard economic or popularity-based metrics in the Japanese context.

\section{Conclusion}
This study proposed nonlinear forecasting models for Japanese lower-house elections using DT and ensemble learning methods. To evaluate whether our models' methodological approach could improve the predictive performance of LBT's pioneering model \cite{lewisbeck2012japanese}, we replicated its substantive theory and dataset. 

All our nonlinear models moderately but consistently outperformed the LBT benchmark model in both in-sample and out-of-sample evaluations. We interpret these findings as offering an alternative methodology to classical linear regression. Our results confirm the validity of LBT's core explanatory variables by demonstrating their predictive power even under a non-parametric framework. At the same time, the consistent improvement in accuracy suggests that relaxing linearity assumptions offers a valuable methodological refinement for capturing the complex dynamics of Japanese elections, albeit with the additional complexity that nonlinear methods entail.

More broadly, this study contributes to the advancement of single-country electoral forecasting. Our models represent one of the earlier applications of nonlinear machine-learning algorithms in this domain. Despite the small sample sizes inherent in country-specific forecasting, our models outperformed linear alternatives, including LBT's, across the majority of elections tested. Future research could adapt and extend this framework to other national contexts by combining our methodological approach with substantive election theory specific to each country, potentially improving forecasting accuracy. Such work, particularly in data-rich environments, would also help address the limitations of a small dataset, confirming the generalizability of the model's hyperparameters.

One of the key goals of LBT's study was to demonstrate that the substantive components of their model---what they referred to as the ``core political economy model''---are cross-nationally applicable. If that is the case, then the methodological components of forecasting models---which are generally less context-dependent---may be even more transferable across borders. We hope our models' methodological approach will serve as a useful reference point for the development of interpretable, country-specific electoral forecasting using machine-learning techniques.

\section*{Data and code availability}
The data, analysis scripts, and source code used to generate the results in this paper are openly available in a GitHub repository at \url{https://github.com/undeadyequ/forecasting_japanese_election_clean}.

\bibliographystyle{unsrt}  
\bibliography{mybib}  

\clearpage 

\appendix
\section{Appendix}
\label{sec:appendix}

\setcounter{table}{0}
\renewcommand{\thetable}{A\arabic{table}}

The following tables present the results of the supplementary evaluation, which was conducted to determine the robustness of the forecasting results of our nonlinear models. Table~\ref{tab:appendix_a1_actual} shows the results of the out-of-sample testing using only the data from 1960--2009, the same period considered in LBT's study. Our DT-gradient boosting model outperformed the LBT benchmark model in terms of \textit{MAE}, which is the primary evaluation criterion. However, the two models yielded identical \textit{RMSE} scores (5.80). 

As discussed in the main text, this parity is driven by the fact that \textit{RMSE} is highly sensitive to statistical outliers \cite{hyndman2006another}, and the DT-gradient boosting model did not forecast the anomalous 2009 election well (see Figure~\ref{fig:fig6}).

Table~\ref{tab:appendix_a2_actual} presents the results of out-of-sample testing, excluding the 2009 election, a possible outlier. The DT-gradient boosting model outperformed the LBT benchmark model.

\begin{table}[htbp]
\centering
\caption{Evaluation of out-of-sample performance on data from 1960 to 2009}
\label{tab:appendix_a1_actual}
\begin{tabular}{lcc}
\toprule
 & \textbf{LBT} & \textbf{DT-gradient boosting} \\
\midrule
\textbf{MAE} & 4.78 & 4.41 \\
 & (1.6) & (2.25) \\
\addlinespace
\textbf{RMSE} & 5.80 & 5.80 \\
 & (1.91) & (3.25) \\
\bottomrule
\multicolumn{3}{l}{\footnotesize Note: Numbers in parentheses represent the standard deviations.}\\
\multicolumn{3}{l}{\footnotesize LBT: Lewis-Beck and Tien; DT: decision tree; MAE: MAE; RMSE: RMSE.}
\end{tabular}
\end{table}
\begin{table}[htbp]
\centering
\caption{Evaluation of out-of-sample performance on data from 1960 to 2017 (excluding 2009)}
\label{tab:appendix_a2_actual}
\begin{tabular}{lcc}
\toprule
 & \textbf{LBT} & \textbf{DT-gradient boosting} \\
\midrule
\textbf{MAE} & 4.56 & 4.08 \\
 & (1.47) & (1.51) \\
\addlinespace
\textbf{RMSE} & 5.33 & 4.79 \\
 & (1.56) & (1.56) \\
\bottomrule
\multicolumn{3}{l}{\footnotesize Note: Numbers in parentheses represent the standard deviation.}\\
\multicolumn{3}{l}{\footnotesize LBT: Lewis-Beck and Tien; DT: decision tree; MAE: MAE; RMSE: RMSE.}
\end{tabular}
\end{table}

\end{document}